\documentclass[12pt]{article}

\usepackage{epsfig}
\usepackage{amssymb}
\usepackage{amsfonts}

\usepackage{color}

\def\be{\begin{equation}}
\def\ee{\end{equation}}
\def\ba{\begin{array}{c}}
\def\ea{\end{array}}

\newcommand{\bea}{\begin{eqnarray}}
\newcommand{\eea}{\end{eqnarray}}

\newtheorem{thm}{Theorem}

\newtheorem{prop}[thm]{Proposition}

\begin{document}


 \begin{center}{\Large \bf

\textcolor{black}{Closed-form construction of exceptional points in a}
two-parametric ${\cal PT}-$symmetric large-matrix model


  }\end{center}

 \begin{center}

\vspace{8mm}

  {\bf Miloslav Znojil} $^{1,2,3,4}$

\end{center}

\vspace{8mm}

  $^{1}$
 Czech Academy of Sciences,
 Nuclear Physics Institute,
 Hlavn\'{\i} 130,
250 68 \v{R}e\v{z}, Czech Republic, {e-mail: znojil@ujf.cas.cz}

 $^{2}$
 Department of Physics, Faculty of
Science, University of Hradec Kr\'{a}lov\'{e}, Rokitansk\'{e}ho 62,
500 03 Hradec Kr\'{a}lov\'{e},
 Czech Republic

$^3$
School for Data Science and Computational Thinking, Stellenbosch
University, 7600 Stellenbosch,
 South Africa

$^4$ Institute of System Science, Durban University of Technology,
{4001}  Durban,
 South Africa


\newpage

\subsection*{Abstract}

Bound states living on a
one-dimensional discrete
$N-$point lattice are
assumed
generated by a
purely imaginary ${\cal PT}$-symmetric potential
in an extreme dynamical regime
near the Kato's exceptional-point singularity.
The feasibility of
the non-numerical localization
of the singularity is achieved, for all $N < \infty$,
via a restriction
of the class of the eligible potentials
to their user-friendlier two-parametric subfamily.

%

\subsection*{Keywords}.

.

discrete imaginary local quantum potentials;

domains of spontaneously unbroken PT-symmetry;

exceptional-point non-Hermitian degeneracies;

corridors of unitary access to quantum phase transitions;

\newpage

\section{Introduction}

The current popularity \cite{BHook} of study
of quantum bound-state systems which are
controlled by a
non-Hermitian but ${\cal PT}-$symmetric local
potential $$V(x) = V^*(-x)$$
has several independent explanations
ranging from their purely mathematical
appeal \cite{book}
and from their intimate connection
with the so-called open (i.e., unstable)
quantum systems
\cite{Ingrid,Nimrod}
up to an experimental
relevance of their multiple non-quantum
analogues, say, in classical
optics \cite{Christodoulides,Rotter}.

In the context of quantum theory
of closed systems itself,
one of the most important
sources of the popularity
of the related
non-Hermitian bound-state Hamiltonians
\textcolor{black}{
 \be
 H=-\frac{d^2}{dx^2}+V(x) \ \neq \ H^\dagger
 \label{schejk}
 \ee
supporting the} real spectra can be seen in \textcolor{black}{a deeply
innovative possibility of a more or less explicit description of a
genuine quantum phase transition (cf., e.g.,
\cite{catast,catastb,preUwe,preSty}).}

\textcolor{black}{Such a possibility has been also analyzed in our
preceding study \cite{I} which will be cited as paper~I in what
follows. }
\textcolor{black}{We emphasized there that
a key role in the phenomenology and applications of Eq.~(\ref{schejk})
is being played by}
the traditional
exceptional-point (EP) singularities  \cite{Kato}
at which a
parameter-dependent Hamiltonian $H=H(g)$
ceases to be
diagonalizable \cite{ali}, i.e., at which it
admits
\textcolor{black}{a realization of a genuine quantum phase transition.}
\textcolor{black}{The essence of the idea is
that in the singular limit of $g \to g^{(EP)}$,
the system passes through}
an abrupt
change
of its physical
probabilistic interpretation
\textcolor{black}{involving at least some of the observables and
including, first of all, the
Hamiltonian itself}
\cite{Heiss,passage}.

\textcolor{black}{A
key to the
construction of the
phase-transition-admitting
potentials
(which would be, in some sense, ``maximally non-Hermitian'',
cf. paper~I)
lies in a reliable
localization
of the EP singularities.}
\textcolor{black}{A suitable method of such a construction
has been found in a discretization
of the differential-operator models (\ref{schejk}).}
\textcolor{black}{In this framework, the real axis of coordinates
$x \in \mathbb{R}$
is being replaced by a finite
and equidistant
$N-$point lattice
of grid points $x_k$ with $k=1,2,\ldots,N$.
For
technical details and for a few related
empirical observations see, e.g., our older paper \cite{ixna3}.}

\textcolor{black}{In such an approach one simplifies the problem
by converting the differential-operator Hamiltonian
of Eq.~(\ref{schejk}) into an $N$ by $N$ matrix
$H^{(N)}= -\triangle^{(N)}+W^{(N)}$
in which the symbol $\triangle^{(N)}$
stands for the standard discrete
one-dimensional Laplacean,
 \be
 \triangle^{(N)}=
 \left[ \begin {array}{ccccc}
  0&-1&0
 &\ldots&0
 \\
 \noalign{\medskip}-1&0&-1&\ddots&\vdots
 \\
 \noalign{\medskip}0&-1&\ddots&\ddots
 &0
 \\
 \noalign{\medskip}\vdots&\ddots&\ddots&0&-1
 \\
 \noalign{\medskip}0&\ldots&0&-1&0
 \end {array} \right]\,
 \label{Ka8pkin}
 \ee}
\textcolor{black}{while $W^{(N)}$ is diagonal matrix with
elements  $W^{(N)}_{kk}=V(x_k)$.}

\textcolor{black}{Another important source of simplification
can be sought and found in the reduction of the
class of the admissible
${\cal PT}-$symmetric
potentials $V(x_k)$ to the mere purely imaginary
functions of the coordinate, with
$V(x_1)=-{\rm i}A=-V(x_N)$,
$V(x_2)=-{\rm i}B=-V(x_{N-1})$,
etc.
At the not too large matrix dimensions $N$,
this makes the study of the sequence}
 \be
  \left[ \begin {array}{cc} -iA&-1\\{}-1&iA\end {array} \right]\,,
 \ \ \ \ \
 \left[ \begin {array}{ccc} -iA&-1&0
 \\{}-1&0&-1\\{}0&-1&iA\end {array} \right]\,,
 \ \ \ \ \ \left[ \begin {array}{cccc} -iA&-1&0&0
 \\{}-1&-iB&-1&0\\{}0&-1&iB&-1\\{}0&0&-1&iA
 \end {array} \right]\,,\ \ldots\,
 \label{models}
 \ee
\textcolor{black}{of the respective Hamiltonian matrices
$H^{(N)}(A,B,C,\ldots)$ tractable by numerical methods and
by the techniques of linear algebra \cite{ixna3}.}

\textcolor{black}{According to paper I, incidentally,
a purely algebraic,
non-numerical construction of the EPs only
remains feasible up to
$N=6$,
i.e., up to
the three-parametric model}
 \be
 H^{(6)}= H^{(6)} (A,B,C)
 =\left[ \begin {array}{cccccc}
  -iA&-1&0&0&0&0
 \\
 {}-1&-iB&-1&0&0&0
 \\
 {}0&-1&-iC&-1&0&0
 \\
 {}0&0&-1&iC&-1&0
 \\
 {}0&0&0&-1&iB&-1
 \\
 {}0&0&0&0&-1&iA
 \end {array} \right]\,.
 \label{threepa}
 \ee
\textcolor{black}{Thus,}
in our present
\textcolor{black}{immediate continuation
of paper I} we
will
re-analyze the project and we will
propose its less ambitious version
which will enable us to
describe
a successful \textcolor{black}{non-numerical,
purely algebraic} solution
\textcolor{black}{of the EP-localization problem
at the larger (i.e., in fact, at
arbitrarily large) finite matrix dimensions $N$.}

\textcolor{black}{In such a reformulation of the project,
our main idea is simple:
In the light of the difficulties as encountered
in the three- and more-parametric models
in sequence
(\ref{models}) (i.e.,
beyond $N=6$),
we will emphasize the need
of having the models with the arbitrarily large
matrix dimensions $N$
(i.e., of
having good approximations of the differential-operator models --
cf., in particular, papers \cite{ixna3,asymptotic} in this respect).}
\textcolor{black}{This means that
we will require that
$C={\rm i}V(x_3)=0$ and
$D={\rm i}V(x_4)=0$, etc.
In other words,
we will only  be interested in the
existence of the EPs supported by
matrices}
\be
 H^{(N)}(A,B)
  =
 \left[ \begin {array}{ccccccc}
 - {\rm i} A&-1&0&0&0&\ldots&0\\
 \noalign{\medskip}-1&- {\rm i} B&-1&0&0&\ldots&0\\
 \noalign{\medskip}0&-1& 0&-1&\ddots&\ddots&\vdots\\
 \noalign{\medskip}0&0&\ddots&\ddots &\ddots&0&0\\
 \noalign{\medskip}\vdots&\ddots&\ddots&-1& 0&-1&0\\
 \noalign{\medskip}0&\ldots&0&0&-1& {\rm i} B&-1\\
  \noalign{\medskip}0&\ldots&0&0&0&-1&  {\rm i} A
  \end {array} \right]\,,
  \ \ \ \ \ N=(2,3,)\,4,5,\ldots\,
 \label{Ka8p2s}
 \ee
\textcolor{black}{i.e., in the search and localization of the
EPs emerging in such a
two-parametric family of the next-to-well-known
quantum bound-state Hamiltonians.}

The presentation of our results will be preceded
by section \ref{seunta} in which we will
\textcolor{black}{introduce a
methodically motivated $B=0$
simplification
of our Eq.~(\ref{Ka8p2s})
yielding the comparatively elementary
but, in
the context of
experimental simulations and
applications in many-particle
physics, extremely useful one-dimensional
boundary-controlled chain model
of the nearest-neighbor-interaction type
considered
in the so-called tight-binding
dynamical regime~\cite{Song,Joglekar}.}

In the next two sections \ref{twoparta} and \ref{twopartbe}
we will present the results concerning the
maximally non-Hermitian
two-parametric models $H^{(N)}(A,B)$ at the respective
even and odd values of
the matrix dimension $N$.
Both of these sections
can be read as an amendment and
completion of the
partial result as published in
our recent paper \cite{asymptotic} in which
we used a brute-force method of a
computer-mediated localization of the EPs,
and in which
such an approach
appeared not to work
beyond $N=13$.
In contrast,
our present use of different method
\textcolor{black}{(based on a redefinition of parameters)}
will be shown to
open the way not only towards
a systematic description of all of the ``long chain''
models with $N \gg 1$
but also towards our understanding of the system in
the continuous-coordinate limit $N \to \infty$.

All of these results will
be summarized in section \ref{summary}
and complemented by a few brief comments in
Appendix A.

\section{\textcolor{black}{One-parametric
Hamiltonians $H^{(N)}(A,B)$ with $B=0$\label{seunta}}}

\textcolor{black}{For introductory
as well as methodical purposes let us}
pay attention to the
simplest possible (viz., just
one-parametric, boundary-controlled)
$N$-by-$N$-matrix forms of our Hamiltonians
\textcolor{black}{of Eq.~(\ref{Ka8p2s}) at $B=0$},
 \be
 H^{(N)}(A,0)=
 \left[ \begin {array}{ccccc}
  -{\rm i} A&-1&0
 &\ldots&0
 \\
 \noalign{\medskip}-1&0&-1&\ddots&\vdots
 \\
 \noalign{\medskip}0&-1&\ddots&\ddots
 &0
 \\
 \noalign{\medskip}\vdots&\ddots&\ddots&0&-1
 \\
 \noalign{\medskip}0&\ldots&0&-1&{\rm i} A
 \end {array} \right]\,.
 \label{Ka8p}
 \ee
In this context we will
have to resolve, first of all,
an older paradox
of an enigmatic structural difference
between  the EP-related quantum
dynamics at  even and  odd $N$s.


\subsection{Boundary-controlled models}

The choice of $B=0$ opens
the way towards several
alternative
physical interpretations of the
non-Hermitian Hamiltonian matrices
(cf., e.g., \cite{chainc}).
One of the most influential applications of the specific
one-parametric
toy-model $H^{(N)}(A,0)$ of Eq.~(\ref{Ka8p})
can be found in the works
by  Jin and Song \cite{Song} and by
Joglekar et al  \cite{Joglekar}. These authors
used such a family of
Hamiltonians
during their study of certain
long-chain quantum systems
in the framework of condensed matter physics.

The analysis
led to a surprising observation that
after the replacement of
$H^{(N)}(A,0)$ by $H^{(N)}(0,B)$ (etc),
i.e., after the loss of
the dominance of the control of boundary via
$A \neq 0$,
the ${\cal PT}-$symmetry becomes fragile
(cf. also \cite{fragile} in this respect).
In the long-chain limit of $N \to \infty$
the ${\cal PT}-$symmetry
even proved to be spontaneously broken.

Unexpected as such a result was,
it
motivated our introduction of the
more general two- and more-parametric problems. In paper I,
in particular,
we tried to mimic
the impurity-caused quantum
phase transitions of condensed-matter physics
as an exceptional-point-mediated phenomenon
in which
an active role of the impurity-representing  gain-and-loss
forces was intended to be
played by {\em all\,}
of the (real) parameters $A, B, C, \dots$ {\em simultaneously}.

Such an enhancement of the flexibility of the model
had to offer a new insight in the
underlying
(i.e., in our eyes, not only the condensed-matter-related)
quantum bound-state
dynamics.
Unfortunately,
such a project appeared feasible
at the first few matrix dimensions $N$ only.
For
a non-numerical analysis at
a general size $N$ of the matrices
it seemed recommendable to return to the
simpler interaction potentials
and, in particular, back to
the models with $B=C=\ldots =0$.

\subsection{Solvability \textcolor{black}{of the $B=C=\ldots=0$ models} via Sturmians}

A few
amendments of the practical
constructive tractability of the
tight-binding
models (\ref{models}) were proposed
in our
two recent papers \cite{asymptotic} and \cite{Burdik}.
Basically, we considered there several subsequences
of our present sets of matrices~(\ref{Ka8p2s})
and (\ref{Ka8p}), respectively.
After such a reduction of the problem we managed
to overcome
several
technical obstacles as encountered in
\textcolor{black}{the genuine
quantum multiparametric-interaction systems}.

First of all we accepted the conclusion that
a key to feasibility
of analytic, non-numerical constructions
should be sought in
a reduction of the number of the variable parameters.
Thus,
after the choice of $B=0$ in Eq.~(\ref{Ka8p2s})
we obtained,
in \cite{Burdik}, the class of
the drastically simplified models (\ref{Ka8p})
in which the single real parameter $A$
acquired the role
of a variable which controls the dynamics.

Having in mind our key mathematical task of
a
localization of the exceptional point
singularities
in the
simplified $N$ by $N$
matrix models
at any matrix-dimension $N$,
we still kept in mind that
even in
such a comparatively
elementary special case
the analysis at the larger $N$s ceases to be easy.
Fortunately, we found an important encouragement in the papers by
Jin and
Song \cite{Song,chainb} who proposed to use
the Bethe ansatz technique, and who
discovered
that the special non-Hermitian
matrix
model (\ref{Ka8p})
is
still
solvable.

The latter authors emphasized that
the manifest non-Hermiticity of such an upgraded
Hamiltonian
with real spectrum
opens the way towards a
phenomenologically innovative
description of the genuine
quantum phase transitions (cf. also \cite{catast}
or \cite{passage}).
From our purely methodical point of view
it was important
that
Jin and Song revealed that
the spectral properties of the model
are not a smooth function of
the matrix dimension $N$.
One must separate the models
with even
$N=2K$
from those with odd
$N = 2K+1$.

\subsection{\textcolor{black}{The even$-N$ construction}}

In \cite{Burdik} we followed the advice and
we
restricted our attention just to the
one-parametric models with even $N=2K$.
Such an additional constraint proved also useful
in our other
applied-mathematics
paper \cite{physcri}
where we used the model to illustrate
the possible role of the EP singularities
in
quantum cosmology.

We emphasized
that
\textcolor{black}{for the family of models (\ref{models})}
the degree of the secular polynomial
grows rather quickly with the growth of $N$
so that at
larger $N$s the
localization of the bound-state
energy roots $E_n=E_n(A)$ becomes a purely numerical task.
At the same time
we noticed that
our secular polynomial
only remains quadratic
when considered as a polynomial
in the variable coupling $A$.
In this manner we could reinterpret the bound-state
energy roots $E_n=E_n(A)$,
at an arbitrarily large $N$,
as a not too complicated closed-form
inverse function
$A=A_n(E)$
called, usually, Sturmian.

An $N=6$ sample of such an inverse function can be found
displayed
as formula Nr.~18 in \cite{physcri}.
In the same paper, in addition,
we introduced
an {\it ad hoc\,} reparametrization
of the
strength of the coupling,
 \be
 A \to A(r)=\sqrt{1-r^2}\,,\ \ \ \ r \in \mathbb{R}
 \,,\ \ \ \ \ N=2K
 \,.
 \label{adhoc}
 \ee
This further simplified the mathematics
and
clarified the structure of the spectrum --
for illustration see also picture Nr. 1 in \cite{physcri}
where, again, $N=6$.

After the reparametrization
of Eq.~(\ref{adhoc}), moreover,
the spectrum has been shown to
stay real at all
of the matrix dimensions $N=2K$ and
at all of the real parameters $r \in \mathbb{R}$.
In a complementary study \cite{emergence}, finally, the latter
Sturmianic solution has been extended to cover
a number of further values of $N$, even or odd
(cf., e.g., the table Nr. 1 in {\it loc. cit.}).
Still,
one of the issues which were
noticed in
\cite{emergence} and \cite{Burdik}
was an enigmatic absence of the odd$-N$ EPs at $r=0$.

\subsection{Odd$-N$ puzzle}


In both of the latter two papers,
no
explanation
\textcolor{black}{of the even/odd $N$ anomaly
has been found}.
In the former paper
\textcolor{black}{the problem has only been circumvented
via}
abandoning the
maximal non-Hermiticity
requirement, and via admitting
that our parameter $A$
can be complex.
Only then, i.e., after an availability of
another, second real parameter
(viz., ${\rm Im}\,A \neq 0$, cf. equation Nr. 19 in {\it loc. cit.}),
a sufficiently
flexible and
satisfactory toy model
has been obtained
describing the emergence
and properties of the
system near its
EP-represented singularities.

In our present continuation of this analysis
we decided to return to the
two-real-parameter model (\ref{Ka8p2s}) living
in a suitably defined
maximally non-Hermitian
dynamical regime.
\textcolor{black}{A decisive enhancement of the insight
in the odd-$N$/even-$N$ puzzle
will be then achieved when}
we set $B=0$
and re-analyze, once more, the
\textcolor{black}{simplified eigenvalue problem involving the}
one-parametric matrices (\ref{Ka8p})
at the odd $N=2K+1$.

\begin{figure}[h]                    
\begin{center}                         
\epsfig{file=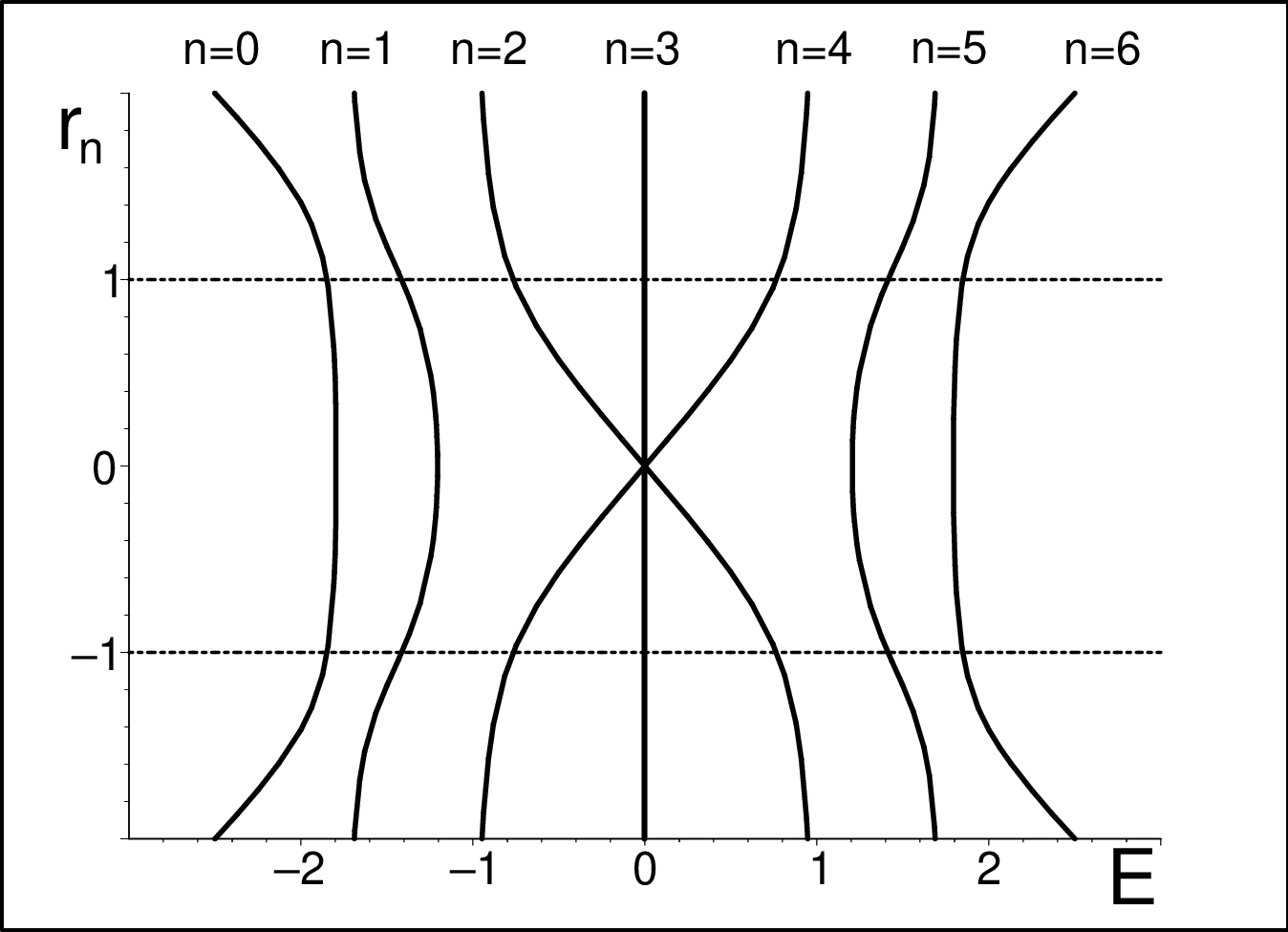,angle=270,width=0.35\textwidth}
\end{center}    
\caption{Sturmian representation of spectrum $r_n(E)$ and
the triple exceptional point of Hamiltonian (\ref{Ka8p})
after reparametrization (\ref{adlihoc})
at $N=7$.
Notice that the model is Hermitian iff $|r|\geq 1$,
quasi-Hermitian iff $0 < |r| < 1$, and unphysical at $r=0$.
 \label{Stglobe}}
\end{figure}

In this setting,
our new and important result
is that
in the constructive
Sturmian-solution context,
the reparametrization formula (\ref{adhoc})
is only appropriate
for the models with even $N=2K$.
In other words,
with
our attention redirected to
the one-parametric models with odd $N$,
the reparametrization of $H^{(N)}(A,0)$
as represented by Eq.~(\ref{adhoc})
must be replaced
by the following different formula,
 \be
 A = A(r)=\sqrt{\frac{K+1}{K}}\,\sqrt{1-\,r^2}\,,
 \ \ \ \ r \in \mathbb{R}
 \,,\ \ \ \ \ N=2K+1
 \,.
 \label{adlihoc}
 \ee
Under this ansatz,
in a way illustrated in Figure \ref{Stglobe},
the system
will fall into
the  EP singularity and
lose its observability
in the limit
of $r \to 0$, i.e.,
along the same lines as it did at the even $N=2K$.

\textcolor{black}{Preliminarily, we may conclude
that, {\it mutatis mutandis},
our experience with the difference
between the changes of variables
(\ref{adhoc}) and (\ref{adlihoc})
should be also transferred, as
a methodical guide, to our
next-step study of the full-fledged $B\neq 0$ models.}

\section{Two-parametric models
with even $N=2K$\label{twoparta}}

\subsection{The task of localization of exceptional points}

For all of the above-listed reasons we
decided to return
to the work with non-vanishing
values of the second real parameter $B \neq 0$.
A partial success in the localization of exceptional points
has been
achieved
in paper
\cite{asymptotic}
in which we
solved the underlying sets of coupled
polynomial equations using
the standard
Gr\"{o}bner's elimination technique.
Unfortunately, we were only successful
at the even Hilbert-space dimensions $N=2K$.
Let us now briefly return to
the methodical message as provided by such a partial
success.

What is essential is that in paper
\cite{asymptotic}
we decided to reduce the
multiparametric  family (\ref{models})
to its
two-parametric subset
(\ref{Ka8p2s}).
After such a rather drastic
reduction of the project,
its original regularization-analysis motivation
was still preserved.
Only
its form and generality have changed.
Even
in the continuous-coordinate-mimicking
$N \to \infty$ limit, the interaction
remained restricted
to a small vicinity of the end-points of the interval of
the coordinate $x \in (-L,L)$.

After the simplification $C=D=\ldots = 0$
we were really able to analyze
the large$-N$
versions of Hamiltonians (\ref{Ka8p2s})
at any even matrix dimension $N = 2K$.
An insight gained during
analysis of the
one-parametric boundary-controlled
models $H^{(N)}(A,0)$
\textcolor{black}{proved essential and}
helped us to overcome the obstacles.

From our present point of view, incidentally,
it makes good sense
to explain
the basic ingredients of the construction
via the first nontrivial choice of $K=2$.
In paper I,
the elementary nature of such a sample model with $N=4$
has been used to shorten the
construction by keeping it purely
numerical
(cf. section Nr. V. A in {\it loc. cit.}).
Now,
with our
methodical
target being the arbitrary$-N$ systems,
the same instructive
model with $N=4$
may and should be dealt with
using
a
more systematic and basically non-numerical
amended approach.

\subsection{Sample construction at $N=4$}


The exact solvability of our Hamiltonian
 \be
 H^{(4)}(A,B)
 =\left[ \begin {array}{cccc} -iA&-1&0&0
 \\{}-1&-iB&-1&0\\{}0&-1&iB&-1\\{}0&0&-1&iA
 \end {array} \right]
 \label{hacty}
 \ee
is,
from
our present methodical point of view,
one of its key merits.
Once we
define the spectrum using secular equation
 \be
 {{\it E}}^{4}+ b\, {{\it E}}^{2}+c=0\,
 \label{eku}
 \ee
with coefficients
 $$
b=b(A,B)= -3+{A}^{2}+{B}^{2} \,,
 \ \ \ c=c(A,B)
 =(1+A\,B)^2-A^2\,,
 $$
all of the four bound-state energies remain obtainable in closed form,
 $$
 E_{\pm,\pm}=\pm 1/2\,\sqrt {6\pm 2\,\sqrt {5-2\,{A}^{2}-6\,{B}^{2}
 +{A}^{4}-2\,{B}^{2}{A}^{2}+{B}^{4}-8\,BA}-2\,{A}^{2}-2\,{B}^{2}}\,.
 $$
This is the result of paper \cite{I}
where we revealed the existence and
completeness of the
non-Hermitian degeneracy of all of the four energy
levels towards their singular and unphysical limit $E^{(EP4)}=0$.

An explicit non-numerical constructive localization
of the degeneracy is a key to the construction of the
Hamiltonians exhibiting the
maximal non-Hermiticity.
Such a localization is
to be
based on the pair of constraints $b(A,B)=c(A,B)=0$
or, in explicit form,
 \be
  -3+{A}^{2}+{B}^{2}=0\,,\ \ \ \ 1+2\,
 BA-{A}^{2}+{B}^{2}{A}^{2}=0\,.
 \label{rucha}
  \ee
In the second item one must have $A \neq 0$
so that one is permitted to eliminate
 $$
 B=\pm 1 -1/A\,.
 $$
This opens the possibility of a simplification
of the procedure
which has not been earlier noticed. Indeed,
after a backward insertion of the latter formula
(i.e., in fact, of the two different formulae)
one arrives at the two
respective polynomial equations for $A=A^{(EP4)}$. The first one
is
 \be
 A^4-2\,A^2- 2\,A+1=0
 \label{minusova}
 \ee
and yields the two real roots, viz., $
0.3715069740$ and $1.683771565$. Its
partner
 \be
 A^4-2\,A^2+ 2\,A+1=0
 \label{plusova}
 \ee
yields the other two real roots $
-0.3715069740$ and $ -1.683771565$.
The
respective
pairs of the quantities $B$ are
 $$
 B=-1.691739510\,,\ \ \ B=0.4060952086
 $$
and
 $$
 B=-0.4060952086\,\ \ \ B=-1.691739510\,.
 $$
We can conclude that both of the
complementary EP4-localizing equations (\ref{minusova})
and (\ref{plusova})
are easy to solve in closed form.

\subsection{Transition matrix}

In paper \cite{asymptotic}, interested readers may find
a straightforward generalization of the latter pair of equations
to all of the even matric dimensions $N = 2K$.
Here, let us only add a rather technical remark
that in the manifestly unphysical EP4 limit
the Hamiltonians cease to be diagonalizable and
acquire a canonical Jordan-block representation
in general \cite{Kato,Heiss}.

In preceding subsection we localized the corresponding EP instant of
the four-fold degeneracy of the eigenvalues,
 \be
 E_{\pm,\pm}^{(EP4)}(A^{(EP4)},B^{(EP4)})=0\,.
 \label{degenrg}
 \ee
What remains to be verified is
the EP4-form of the
non-diagonalizability
of the singular Hamiltonian.
This means that we have to
verify
that
its
canonical form
really is
just the
four-by-four Jordan matrix,
i.e., that there exists
the so called transition matrix $Q^{(4)}$
such that
 \be
 H^{(4)}(A^{(EP4)},B^{(EP4)})\,Q^{(4)}
 =Q^{(4)}\, \left[ \begin {array}{cccc}   0&1
 &0& 0
 \\{}
 0& 0&1
 &0
 \\{}
 0& 0&0&1
 \\{}
 0& 0& 0& 0\end {array} \right]\,.
 \label{tranzi}
 \ee
\textcolor{black}{In paper I it has been shown
that at $N\geq 6$}
the set of the parameters admitting the EP degeneracy
ceases to be unique.
\textcolor{black}{Marginally, let us add that in paper I,
the explicit and constructive
proof of the existence of the ambiguity
has only been made available at $N=6$
and after the choice of $B\neq 0$ and $C \neq 0$. Still,}
only one of the solutions
has been presented. It was
characterized by the largest parameters
$A^{(EP6)}\approx 2.05$, $B^{(EP6)}\approx 0.86$
and $C^{(EP6)}\approx 0.26$ (cf. Table Nr. I in {\it loc. cit.}).
The other solution --
with $A^{(EP6)}\approx 1.15$, $B^{(EP6)}\approx 0.43$
and $C^{(EP6)}\approx 0.23$ --
has been discarded as
asymptotically subdominant, i.e., not
maximally non-Hermitian.

In {\it loc. cit.} we also
proposed to suppress the ambiguity
by
\textcolor{black}{an additional requirement
mimicking a certain preferred} asymptotic growth
of some conventional imaginary continuous-coordinate potentials
at $|x| \to \infty$.
We decided to admit only the
non-oscillatory shapes
of the functions $V(x)$
at finite $N < \infty$.
In this manner we
excluded all of the
EP4-degeneracies with negative $B<0$.

This \textcolor{black}{convention} makes also the coordinates
of the illustrative $N=4$ degeneracy
unique,
 $$
 A^{(EP4)}= 1.683771565\,,\ \ \ \
  B^{(EP4)}= 0.4060952085\,.
 $$
What is now
\textcolor{black}{only missing}
is the proof of the
simultaneous confluence of
all of the four respective eigenvectors
or, equivalently \cite{Kato},
of the existence of the
transition-matrix solution $Q^{(4)}$
of Eq.~(\ref{tranzi}).
As long as $N=4$ is not large, the latter
linear-algebraic proof is easy to deliver.
In fact, the contruction yields the result
 $$
 Q^{(4)}= \left[ \begin {array}{cccc}   i&-
 1.835086683& - 1.683771565\,i& 1
 \\{} 1.683771562&
 2.089866772\,i&- 1& 0\\{}
 - 1.683771565\,i& 1& 0& 0
 \\{}- 1& 0& 0& 0\end {array} \right]\,.
  $$
This confirms that
the exceptional point is of the maximal
(i.e.,
of the fourth) order.

In principle, a similar demonstration of the
loss of the diagonalizability
and of the non-Hermitian EPM degeneracy
(with a suitable degree of degeneracy $M$,
presumably equal to four)
has to be performed
at any even
matrix dimension $N=2K$. In practice, such a proof remains feasible
even at the larger $K$s, but
in such a case the solution of an analogue of
Eq.~(\ref{tranzi}) becomes numerical and
requires
a careful control of the influence of the round-off errors.
An illustrative sample of such a check (using
the floating-point arithmetics with the variable number of digits)
may be found described, e.g., in paper \cite{4and5}.

\section{Two-parametric models
with odd $N=2K+1$\label{twopartbe}}

Models (\ref{Ka8p2s}) with the odd
matrix-dimensions $N=2K+1$
appeared to be much
less
user-friendly and
suitable
for a constructive analysis
of the system
in its maximally non-Hermitian
dynamical regime.
Even an ample use of the computer-assisted symbolic manipulations
did not lead to the results
which would be compact and comparable with those obtained, in
\cite{asymptotic},
at the even $N$s.

At the odd $N$s,
we
were able to
overcome the emerging complications and to
evaluate the
eligible sets of the critical
EPM parameters only at $N\leq 13$.
Thus, the vicinity of the
boundaries of the physical parametric domains
${\cal D}_{physical}^{(2K+1)}$
supporting the existence of
the maximally non-Hermitian Hamiltonians
$H^{(2K+1)}(A,B)$ was only made known
at the first few $K$s.

The
\textcolor{black}{restriction to  $N\leq 13$}
was caused by an inadequacy of the method.
In what follows we intend to
show that after an appropriate amendment of the method,
also the models (\ref{Ka8p2s}) with odd $N=2K+1$
can be re-classified as solvable.
Our understanding and
description of the maximally non-Hermitian
forms of Hamiltonians $H^{(N)}(A,B)$
will be made complete.

\subsection{The Gr\"{o}bner-basis construction}

In contrast to the success of the
construction
at the even $N=2K$,
its odd$-N$ analogue
behaved differently.
Our computer-assisted
search for the EP singularities
only worked up to $K\approx 6$
in
\cite{asymptotic}
(see the last Table and the last Figure in {\it loc. cit.}).
Our symbolic-manipulation-mediated
localization of the EP extremes suddenly
turned inefficient.

An intuitive insight in
the
$K-$dependence of the results
ceased to be available as early as
at $K=4$
(cf. equation Nr. 22
in {\it loc. cit.}).
The only good news was that
in spite of
the loss of insight,
the more or less routine
identifications of the critical exceptional-point parameters
with the roots of a computer-generated
effective polynomial
still did work.

\textcolor{black}{In particular,}
the odd$-N$ secular polynomial
 $$
  P^{(2K+1)}_{(A,B)}(E)=
  \det \left [\,H^{(2K+1)}(A,B)-E \,\right ]
  $$
was still available in an unessentially modified form
  \be
 -E^{2K+1} + c_1(A,B)
 E^{2K-1}+ c_2(A,B)
 E^{2K-3}+
 \ldots
 + c_{K-1}(A,B)E^{3}+c_{K}(A,B)E^{}\,.
 \label{mikula}
 \ee
For the purposes of the localization
of a non-Hermitian degeneracy
 \be
 \lim_{\{A,B\} \to \{A^{(EPM)},B^{(EPM)}\} } E^{(n)}(A,B)=
 0\,,\ \ \ n=n_1\,, n_2\,,\ldots\,,n_M
 \label{eld}
 \ee
of the energy levels
(with, typically, $M=3$), therefore,
we could still follow the methodical guidance as provided
by the $B=0$ simplification
(cf. Figure \ref{Stglobe} above).
Along these lines, at the first few $K$s at least,
we could still
evaluate the
odd$-N$ coefficients of the
polynomials as sampled in Table \ref{pexp3b}.

\begin{table}[h]
\caption{Relevant coefficients in secular polynomial
(\ref{mikula}) at odd $N=2K+1$.}
 \label{pexp3b}
\begin{center}
\begin{tabular}{||c||c|c||}
\hline \hline
  $K$ & $ c_K(A,B)$& $ c_{K-1}(A,B)$
  \\
  \hline \hline
  2&$-3-2\,BA+2\,{A}^{2}-{B}^{2}{A}^{2}$&$4-{A}^{2}-{B}^{2}
  $\\
  3&$ 4+4\,BA-3\,{A}^{2}+2\,{B}^{2}{A}^{2}$&$
  -10+2\,{B}^{2}-2\,BA+4\,{A}^{2}-{B}^{2}{A}^{2}
  $\\
 4 &$-5+4\,{A}^{2}-6\,BA-3\,{B}^{2}{A}^{2} $&$
 20-3\,{B}^{2}+8\,BA-10\,{A}^{2}+4\,{B}^{2}{A}^{2}
  $\\
 5 &$ 6-5\,{A}^{2}+8\,BA+4\,{B}^{2}{A}^{2}$&$
 -35+20\,{A}^{2}-20\,BA+4\,{B}^{2}-10\,{B}^{2}{A}^{2}
  $\\
 \vdots & \vdots &\vdots \\
\hline \hline
\end{tabular}
\end{center}
\end{table}

On these grounds we recalled the successful even$-N$ procedure and
postulated again the validity of
the two coupled algebraic equations
 \be
 c_K(A^{(EP3)},B^{(EP3)})=0\,,\ \ \
 c_{K-1}(A^{(EP3)},B^{(EP3)})=0\,.
 \label{elda}
 \ee
Thus,  we were able to obtain
all of the eligible singularity coordinates
as the real computer-provided roots
of such a set, in principle at least.

In practice, the results of these calculations were not satisfactory.
First of all, the $K-$dependence
of the root-determining quartic polynomials
were utterly enigmatic
(compare, e.g., the two Figures Nr. 5 and 6
in {\it loc. cit.}).
Secondly,
the formulae representing the computer-generated
parameters $A^{(EP3)}$ and $B^{(EP3)}$, albeit exact,
were too long for a printed display.
Thirdly,
even the numerical
determination of these quantities
required a different algorithm at $K\leq 4$
and at $K>4$. At $K=2$, $K=3$ and $K=4$, indeed,
we were still able to obtain
the value of $A^{(EP3)}$
as one of the roots of a comparatively elementary polynomial
(cf. the respective equations Nr. 20, 21 and 22 in {\it loc cit.}).
At the next two values of $K=5$ and $K=6$, nevertheless,
the algorithm suddenly changed the format of solutions and
produced, instead, the
critical values of $B^{(EP3)}$
(cf. the respective two Tables Nr. 3 and 4
in {\it loc. cit.}).

The enigma is to be resolved in what follows.

\subsection{Change of variables}

%

After an elementary redefinition
 $$
 c_K \to d_k=(-1)^{K+1}\,c_K\,,\ \ \ \ c_{K-1} \to
 d_{K-1}=(-1)^{K}\,c_{K-1}
 $$
of the coefficients in the secular polynomial of Eq.~(\ref{mikula})
we may
introduce an auxiliary (i.e., redundant) real parameter $X=1+AB$.
This enables us to
replace Table~\ref{pexp3b}
by its more compact alternative, viz., by Table~\ref{pexp3x}.

\begin{table}[h]
\caption{Simplified polynomials.
 } \label{pexp3x}
\begin{center}
\begin{tabular}{||c||c|c||}
\hline \hline
  $K$ & $ d_K(X,A)$& $ d_{K-1}(X,A)$
  \\
  \hline \hline
  2&$2+X^2-2\,{A}^{2}$&$4-{B}^{2}-{A}^{2}
  $\\
  3&$2+2\,X^2-3\,{A}^{2}$&$
  9+X^2-2\,{B}^{2}-4\,{A}^{2}
  $\\
 4 &$2+3\,X^2-4\,{A}^{2}$&$
 16+4\,X^2-3\,{B}^{2}-10\,{A}^{2}
  $\\
 \vdots & \vdots &\vdots
  \\
\hline \hline
\end{tabular}
\end{center}
\end{table}


The first column of Table \ref{pexp3x}
can be now extrapolated yielding
the general formula
 \be
 d_K=2+(K-1)\,X^2-K\,A^2\,,\ \ \ \ \  K = 2, 3, \ldots
 \,.
 \label{easyel}
 \ee
Using abbreviation $Q_m=m(m-1)(m-2)/6$,
the second column of Table \ref{pexp3x}
can be
extrapolated to all $K$s as well,
 \be
 d_{K-1}=K^2+Q_K\,X^2-(K-1) \,B^2-Q_{K+1}\,A^2\,,
 \ \ \ \ \ K = 2, 3, \ldots\,.
 \label{afinequ}
 \ee
The two terms with
$Q_K$ and $Q_{K+1}$ dominate at large $K\gg 1$
since $Q_m = m^3/6 +{\cal O}(m^2)$.

In the next step, the apparently redundant auxiliary
variable $X=1+AB$ is to be reinterpreted as
a substitute for $B=B(X,A)=(X-1)/A$
(in our present study we are never going to admit $A=0$).
Thus, having, in this manner,
the modified
explicit representation of $d_K=d_K(X,A)$ and
$d_{K-1}=d_{K-1}(X,A)$,
we are now prepared
to search for the
possibility of the
EPM degeneracy with $E_n \to E^{(EPM)}=0$
at some energy-level subscripts $n$.

The ``maximal'' (i.e., five-level, EP=EP5 \cite{I}) form of
the degeneracy
may be encountered when we guarantee that
 \be
 d_K(X,A)=0\,,\ \ \ \ d_{K-1}(X,A)=0\,.
 \label{sikula}
 \ee
This means that we have
to find
all of the admissible (i.e., real) EP5 roots
$X=X^{(EP5)}$ and $A=A^{(EP5)}$
of such a coupled
pair of polynomial algebraic equations, since
a subsequent ultimate elementary reconstruction
of $B=B^{(EP5)}=(X^{(EP5)}-1)/A^{(EP5)}$ remains easy.

The solution of system (\ref{sikula})
can proceed via
the elimination of the second unknown quantity $A^{(EP5)}$
from the first item of Eq.~(\ref{sikula}).
Once we rescale
 $$
 A^{(EP5)} \to Y^{(EP5)}=[{A}^{(EP5)}]^{2}\,K\,,
 $$
we obtain the
expression
 $$
 Y^{(EP5)}=Y^{(EP5)}(K,X)
 =2+(K-1)\,X^2\,,
 $$
i.e., an elementary $K-$dependent function of $X$.

It remains for us to recall the
second EP5
constraint, i.e., requirement $d_{K-1}=0$
or, equivalently, $K^2\,A^2\,d_{K-1}=0$
or, optimally, $K\,Y\,d_{K-1}=0$.
Thus, after we
insert formula (\ref{afinequ}),
we end up with relation
 $$
 K^3\,Y(K,X)+K\,Y(K,X)\,Q_K\,X^2
 -\,Q_{K+1}\,Y^2(K,X)-K^2\,(K-1) A^2\,B^2=0
 $$
where we only have to substitute the expression
$(X-1)^2$
for the product $A^2B^2=(X-1)^2$
in the last term.
This completes the proof of our following main result.

\begin{prop}. All of the eligible EP5 parameters
$X=X^{(EP5)}$
are
defined as real roots of the quartic (i.e., exactly solvable) polynomial
 \be
 Z(K,X)=K^3\,Y(K,X)+K\,Y(K,X)\,Q_K\,X^2
 -\,Q_{K+1}\,Y^2(K,X)-K^2\,(K-1) (X-1)^2\,
 \label{nereska}
  \ee
at all $K = 2, 3, \ldots$.
\end{prop}

\begin{figure}[h]                    
\begin{center}                         
\epsfig{file=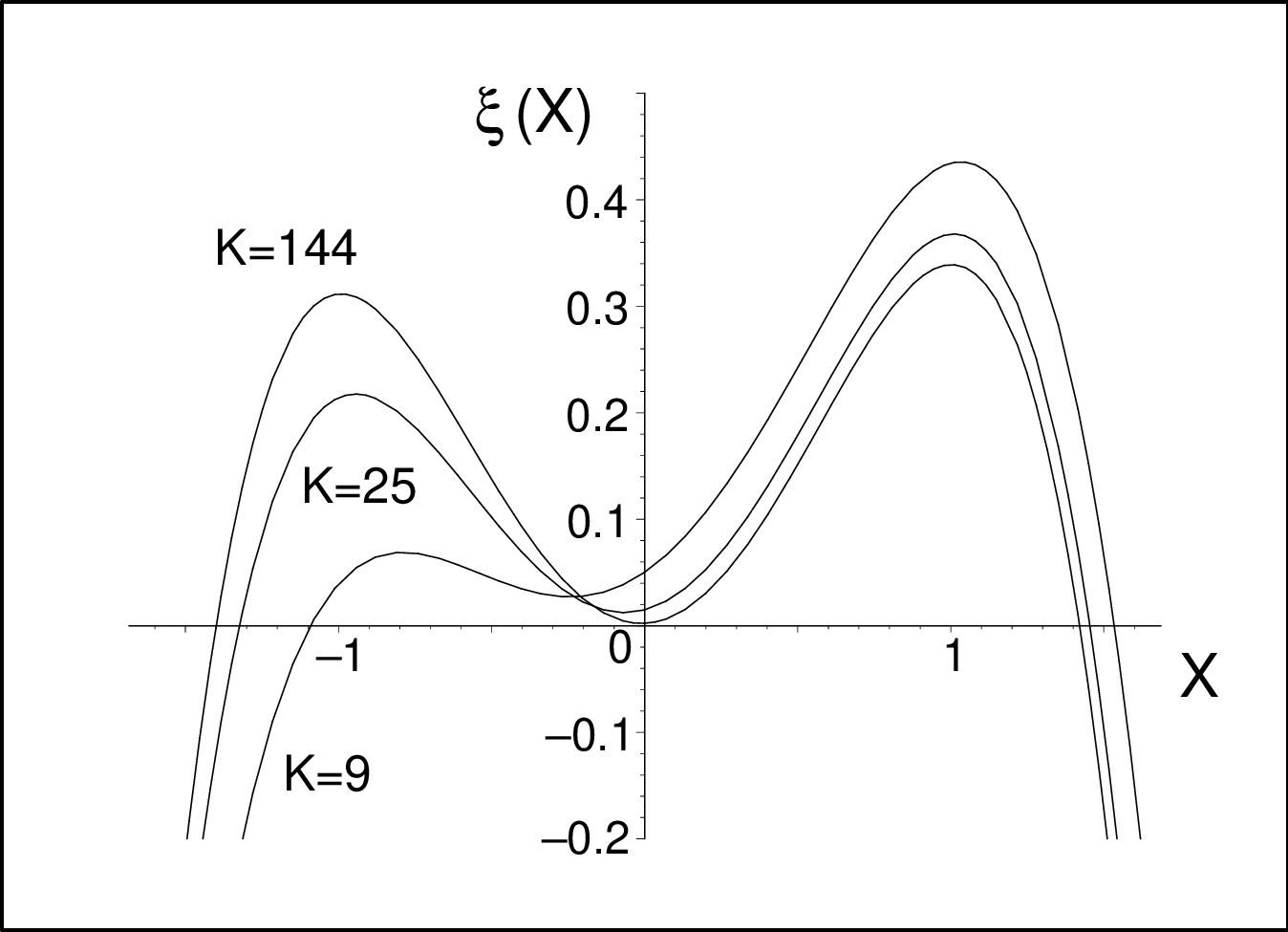,angle=270,width=0.35\textwidth}
\end{center}    
\caption{Rescaled polynomials $\xi(X)=Z(K,X)/K^4$
exhibiting a monotonous asymptotic convergence
of their two real roots $X^{(EP5)}_{\pm}(K)$
towards
$X^{(EP5)}_{\pm}(\infty)= \pm \sqrt{2}$.
 \label{kglobe}}
\end{figure}

\subsection{Large$-K$ simplification}

For the purposes of a suitable approximation of the
roots $X=X(K)$ of
the functions $Z(K,X)$ of Eq.~(\ref{nereska})
in the large$-K$ domain,
it makes sense to
replace the exact polynomial $Z(K,X)$ by its
rescaled approximate
version
 $$
 3\,Z(K,X)/K^4=
 2\,{X}^{2}-{X}^{4}
 +(1+6\,X-7\,{X}^{2}+5\,{X}^{4}/2)/K + {\cal O}(1/K^2)\,.
 $$
In a way confirmed also by a graphical representation
of the exact function of $X$
in Figure~\ref{kglobe}, such an
asymptotic simplification
of the polynomial immediately reveals that
in the limit of $K \to \infty$ we get the quadruplet of
asymptotic
roots $\{-\sqrt{2},0,0,\sqrt{2}\}$.

At finite $K$,
the first-order correction ${\cal O}(1/K)$
removes the degeneracy of
the two smaller roots (cf. the center of Figure \ref{kglobe}).
It shifts them both
off the real axis.
Thus,
at any sufficiently large but finite $K$,
these roots cannot be real.

Concerning the two
larger (and safely real) roots,
the search for
their optimal numerical
as well as non-numerical
approximate representation
remains straightforward.
For the purpose
we may use  ansatz
 $$
 X=X_\pm = \pm \sqrt{2} + Y_\pm^{(1)} /K + Y_\pm^{(2)} /K^2
 + \ldots\,.
 $$
Elementary insertion yields the
subdominant correction
 $$
 Y_\pm^{(1)} =\frac{3}{2}\mp \frac{3}{8}\sqrt{2}\,.
 $$
In a way re-confirming
the graphs of Figure \ref{kglobe}, both of these two
corrections are positive.
Similarly,
after a slightly longer calculation
the next-term coefficient can be deduced to be
 $$
 Y_\pm^{(2)} =\mp \frac{189}{128}\sqrt{2}+1\,
 $$
(i.e., it already has
both signs), etc.

\section{Summary\label{summary}}

The most
remarkable results of our present study
of the specific
two-parametric Hamiltonians $H^{(N)}(A^{},B^{})$
can be seen
not only in a
completion of the project of paper I
but also
in an innovative
resolution of some of the related technical
problems.
One of the most serious ones has been formulated
in our recent paper
\cite{asymptotic} (reporting the results of
extensive computer-assisted
calculations) in which
we \textcolor{black}{felt forced to conclude}
that even for the
two-parametric subset \textcolor{black}{(\ref{Ka8p2s})}
of the $N$ by $N$ matrix models
$H^{(N)}(A^{},B^{})$,
there
emerges a
deep difference
between the
Hamiltonians with
even $N=2K$
(which have been found solvable at all $K$s)
and the
Hamiltonians with odd
$N=2K+1$
(which only have been found to be
solvable at the first few smallest $K$s).

In our present paper we managed to \textcolor{black}{overcome and}
eliminate
such a counterintuitive separation.
Now we only have to add that such a
completion
of the  project of \textcolor{black}{construction of
the discrete but maximally non-Hermitian potentials
under the constraint of an unbroken {\cal PT} symmetry} has
several immediate conceptual as well as
phenomenological consequences.

Briefly, one can say that
even in the maximally non-Hermitian dynamical regime
the description of the
evolution of the closed quantum systems
can and should be perceived as standard, i.e.,
fully compatible with the conventional
probabilistic quantum theory.
Indeed, once we stay
in the unbroken-${\cal PT}-$symmetry dynamical regime
(i.e., once the parameters
stay inside a
\textcolor{black}{corridor
formed by a}
spectral-reality-supporting domain ${\cal D}$ called ``physical''
\cite{corridors}),
we become able to
employ one of the correct
physical inner-product metrics
\textcolor{black}{(cf. Appendix A below)}
allowing us to conclude that
the bound-state-energy spectrum remains real, discrete and
bounded from below.

It makes sense to emphasize that we only offered here
the localization
of the EP singularities \textcolor{black}{for the mere two-parametric
potentials.}
Even then,
the construction of what we call maximally non-Hermitian
models
\textcolor{black}{
is fairly nontrivial in general (cf. also
a brief comment \cite{Burdik}
in this respect), mainly because it
requires} the evaluation of roots
of an auxiliary polynomial of the degree
\textcolor{black}{which grows quickly with
the number of parameters}.
\textcolor{black}
{Thus,} we are forced
to admit
that
\textcolor{black}{whenever the number of parameters exceeds two,}
``any extrapolation of our constructions \ldots
to a hypothetical continuous-coordinate limit $N \to \infty$ \ldots
does not seem to be feasible'' \cite{I}.

We found that
\textcolor{black}{it makes sense to restrict the
number of the variable parameters to two. Then,}
the difficulties can be circumvented via
a separate treatment of the
models with the even and
odd $N$s.
\textcolor{black}{As a consequence,
we were able to show that
the models with two parameters appeared solvable at all $N$.}
It became also clear
that
such a class of models is still rich enough.
Under the control by the two-parametric Hamiltonians
the quantum system is still flexible enough to
evolve towards an EP-mediated
phase transition.

In the context of mathematics,
a key merit of our models is their exact solvability.
In particular, the localization
of the EP singularity has ultimately been found non-numerical
at all $N<\infty$,
\textcolor{black}{irrespectively of their parity}.
Such a success
was due to a lucky coincidence
of the  well known
necessity of an exact localization of the EP
coordinates $A^{(EP)}$ and $B^{(EP)}$
(reflecting the sensitivity of the degeneracy
to the numerical
round-off errors which is truly enormous \cite{4and5})
with the
(not quite expected) simplicity
of the EP-determining polynomial equations
which appeared to be of the fourth order,
i.e.
which were solvable exactly at all
\textcolor{black}{of the Hilbert-space dimensions}
$N$.

\newpage


\newpage

\section*{Appendix A. The concept of hidden Hermiticity}

\subsection*{A. 1. Dyson maps}

Freeman Dyson
was probably the first physicist
who revealed that
the description
of the bound-state spectrum
of a unitary quantum system
need not necessarily be based on
a Hamiltonian which is
self-adjoint,
$\mathfrak{h}=\mathfrak{h}^\dagger$.
A key motivation to his discovery (cf. \cite{Dyson})
was that
for a many-body system
of his interest,
all of
the conventional calculation
strategies appeared
inefficient.
He also found the
reason:
None of
the standard auxiliary unitary isospectral transformations
 \be
 \mathfrak{h}\to \mathfrak{h}'
 ={\cal U}\,\mathfrak{h}\,{\cal U}^{-1}\,,
 \ \ \ \ \ \ {\cal U}^{-1}={\cal U}^\dagger
 \label{map}
 \ee
(opening the way towards
an ultimate diagonalization)
incorporated,
due to the over-restrictive
unitarity constraint,
the most essential features
of the underlying quantum dynamics.

Dyson
solved the puzzle:
He
removed the unitarity constraint and
replaced ${\cal U}$ by an entirely general
(i.e., more flexible)
invertible operator
$\Omega$ \cite{Twin}.
In this manner he
extended the class of
the available simplifications yielding
a more general isospectral
representation $H$ of the Hamiltonian,
 \be
 \mathfrak{h}\to H
 =\Omega\,\mathfrak{h}\,\Omega^{-1}\,,
 \ \ \ \ \ \ \Omega^\dagger\Omega=\Theta\neq I\,.
 \label{Dymap}
 \ee
In his particular study of a system
of interacting fermions he could, due to
the manifest non-unitarity of $\Omega$,
incorporate the information about
an -- empirically well known --
excessive strength of the particle-particle
correlations
directly into the structure of $\Omega$.
As a consequence, he could move from
the divergent variational
calculations based on $\mathfrak{h}$
to the quickly convergent variational
calculations based on $H$
(consult also review paper \cite{Geyer}
for a
deeper analysis of a few related technical details).

Due to the possibility of generalization
$I \to \Theta=\Omega^\dagger\Omega \neq I$ there emerged
several
innovations
concerning both
the quantum theory
and its applications \cite{Mousse}.
It is now widely accepted
that a quantum system can be interpreted as
unitary even when
 \be
 H\neq H^\dagger= \Theta\,H\,\Theta^{-1}
 ~\label{prvdi}
 \ee
\textcolor{black}{provided only that}
we find a suitable
Hermitian and positive-definite matrix $\Theta$.
\textcolor{black}{In this manner,} the
conventional condition of the Hermiticity of the
Hamiltonian is relaxed and replaced by the so called
quasi-Hermiticity
(cf., e.g., comprehensive reviews \cite{book,ali,Geyer}
for details).

One of the most remarkable
rediscoveries
of these ideas
is due to
Bender with Boettcher \cite{BB} who
turned attention
to the most elementary
differential-operator
Hamiltonians \textcolor{black}{of Eq.~(\ref{schejk})}. They
conjectured
that
the evolution can be required
unitary even when
the potential itself is not real
(cf. also review~\cite{Carl}).

\subsection*{\textcolor{black}{A. 2. Discrete Schr\"{o}dinger equations}}

Our interest in
${\cal PT}-$symmetric
differential-operator
bound-state
models
dates back
to 1998
(cf. \cite{Rafa}) and to
the
Bender's and Boettcher's
letter
\cite{BB}
in which these authors
turned attention to ordinary differential
Schr\"{o}dinger
equations
 \be
 -\frac{d^2}{dx^2}\psi_n(x) +V(x)\psi_n(x)=E_n\psi_n(x)\,,
 \ \ \ \ \ x\in \mathbb{R}\,,
 \ \ \ \ \ \psi_n(x) \in L^2(\mathbb{R})\,,
 \ \ \ \ \ n = 0, 1, \ldots\,
 \label{se}
 \ee
\textcolor{black}{containing the complex but
${\cal PT}-$symmetric
potentials.}
In spite of a highly unusual choice of these specific local potentials,
the spectra were found --  also in many other papers --
real and non-degenerate, i.e., in principle, observable
\cite{usulumb,usulumc,Ema}.

\textcolor{black}{One of the most efficient
methods of dealing with the differential-operator
eigenvalue problems of the form of Eq.~(\ref{se})
is their discretization.}
In the existing related literature,
many relevant results may be found for the
special discrete models with a variable
matrix-dimension $N$
(cf., e.g., \cite{Reboiro}).
Let us recall now
some of their overall features, keeping in mind that
our own choice of the mathematical
simplifications was to
restrict the control of the dynamics to
a small vicinity of
the boundary of the interval of coordinates.
%

A key motivation of our study of such a class of matrix toy-models
is fourfold. Firstly,
many
$N$ by $N$
matrix
models may remain useful,
for the purely methodical purposes,
even far beyond their most popular
samples using $N=2$.
So we filled the gap and showed that
also the  analysis of a few of
their immediate
$N>2$
descendants may
remain feasible as well as instructive.

Secondly, we emphasized that there exist
multiple alternative
families of $N$ by $N$ matrix models which
are all solvable and suitable
for phenomenological purposes.
We pointed out that
every such a solvable model
has to be assigned a domain
${\cal D}={\cal D}^{(N)}={\cal D}^{(N)}_{\rm physical}$
of its parameters for which the spectrum
of the energies remains
real and non-degenerate,
i.e., admitting a conventional bound-state interpretation.

Thirdly,
for the most general sequence of models (\ref{models})
characterized by as many as $[N/2]$ real parameters $A, B, \ldots$
we were not able to answer, in a fully explicit and
constructive manner, the
question of
the existence
of a Hilbert space ${\cal H}_{\rm physical}^{(N)}$
in which our Hamiltonians $H^{(N)}(A, B, \ldots)$ would generate
the standard (i.e., unitary and stable) evolution of the system.
\textcolor{black}{For this reason, we decided to restrict
the number of these parameters to two}.

Fourthly, we imagined that a
true physical appeal and relevance of the
manifest non-Hermiticity
of every separate model $H^{(N)}(A, B)$
will increase with the decrease of the distance
of the set of its parameters from the
boundary $\partial {\cal D}$
of the ``physical''
domain of the unitarity-supporting parameters.
In such a ``maximally non-Hermitian'' dynamical regime,
obviously, one can expect the
most exciting deviations from the
predictions
provided
by the conventional Hermitian models.

\end{document}